# Title: "Observation of exotic water in hydrophilic nanospace of porous coordination polymers"

Authors: Tomoaki Ichii[1*], Takashi Arikawa[1], Kenichiro Omoto[2], Nobuhiko Hosono[2], Hiroshi Sato[2], Susumu Kitagawa[2], Koichiro Tanaka[1,2*]

**Affiliations:**

[1]Department of Physics, Graduate School of Science, Kyoto University, Sakyo-ku, Kyoto 606-8502, Japan.

[2]Institute for Integrated Cell-Material Sciences, Kyoto University, Sakyo-ku, Kyoto 606-8501, Japan.

*Corresponding to: kochan@scphys.kyoto-u.ac.jp (K.T.); ichii.tomoaki.46v@st.kyoto-u.ac.jp (T.I.)

**Abstract:**

The fundamental understanding of water confined in porous coordination polymers (PCPs) is significantly important not only for their applications such as gas storage and separation, but also for exploring the confinement effects in the nanoscale spaces. Here, we report the observation of an exotic water in the well-designed hydrophilic nanopores of PCPs. Single-crystal X-ray diffraction found that nanoconfined water has an ordered structure that is characteristic in ices, but infrared spectroscopy revealed a significant number of broken hydrogen bonds that is characteristic in liquids. We found that their structural properties are quite similar to those of ***solid-liquid supercritical water*** predicted in hydrophobic nanospace at extremely high pressure. Our results will open up not only new potential applications of exotic water in PCPs to control chemical reactions but also experimental systems to clarify the existence of solid-liquid critical points.

**Main Text:**

Porous coordination polymers (PCPs) or Metal-organic frameworks (MOFs)[1] have emerged as a new class of nanoporous materials constructed from organic ligands and metal ions. Compared with other porous materials, PCPs have an excellent ability to control pore structures, sizes, and surface properties, as a result of the wide variety of combinations of organic ligands and metal ions. This capability means that the nanopores of PCPs could be designed to realize an exotic state of adsorbed gas molecules not seen in the bulk phase using confinement effects[2]. Such state is interesting for molecular physics in nanospaces and also new applications to facilitate chemical reactions in nanopores of PCPs.

Among gas molecules, it is known that water molecules are significantly affected by the size, geometry, and inner wall properties of the confined space. The previous studies show that confinement effects occur in spaces narrower than 100 nm scale and can be classified into two regimes. The first regime appears in spaces from 2 nm to a few tens of nanometers, where the confined water is conceptually divided into a water core with bulk liquid-like properties and a shell of water existing around the inner walls with properties different from bulk water [3,4]. The local water around the inner walls and its electrostatic forces affect the collective behavior of bulk liquid-like water, resulting in a higher viscosity and proton mobility than bulk water [5,6]. The second regime appears in spaces narrower than 2 nm, where all water molecules interact with inner walls [7–9]. In this regime, nanoconfined water is strongly affected by the space restrictions and overlapped potentials of the inner-walls, giving rise to exotic properties. Among nanoporous materials, CNTs have been extensively used as a well-defined hydrophobic nanospace[8,10,11]. Previous studies showed that the H-bonds of water in hydrophobic nanospace are highly restricted, resulting that confined water at ambient temperature and pressure has a number of H-bonds in between gas and liquid water. The intermediate value is a signature for liquid-gas supercritical water[8]. Furthermore, recent molecular dynamics simulations predict an intriguing possibility of hydrophobic confinement to achieve solid-liquid supercritical water[12] when water density becomes comparable to liquid water (ex. the critical density = 1.33 g/cm$^3$ in nanofilms [13]). Due to the space limitations, the structural symmetry is not a well-defined order parameter in nanospaces. This allows the existence of a solid liquid critical point, which is generally restricted in the bulk phase by the famous symmetry argument that an isotropic liquid cannot continuously transform into a solid with a discrete symmetry[14,15]. Above a critical point, water is allowed to have intermediate physical parameters (ex. number of H-bonds and diffusivity) between the limits of ice and liquid phase.

Considering the previous studies, water molecules confined in the 1 nm-size nanospace with the large density have a possibility to exhibit an exotic state not seen in the bulk phase. In this paper, we realized confined water with a density of 1g/cm$^3$ in the 1 nm-size hydrophilic nanochannels of PCPs, which provides an attractive potential for water molecules. The potential can dramatically lower the free energy of the adsorbed state due to the large gain of the enthalpy and increase the density of confined water. We investigated the structural and dynamical properties of the confined water in PCPs. By using X-ray diffraction analysis and IR spectroscopy, we revealed that water in PCPs has an ordered structure (characteristic in ices) with a significant number of broken H-bonds (characteristic in liquids). We showed that nanoconfined water in **PCP-1** is exotic with no corresponding states in the bulk phase diagram. We found that the structural properties of water in **PCP-1** are quite similar to those of ***solid-liquid supercritical water*** in CNTs. This result will open up new potential applications of exotic water in PCPs to control chemical reactions but also experimental systems to clarify the existence of solid-liquid critical points.

**Results**
**Isotherm of water adsorbing PCP.** We synthesized a water-adsorbing PCP that is composed of copper ions and isophthalates. The water adsorption/desorption isotherm was measured at 298 K, as shown in Fig. 1A. At $P/P_0 = 0.04$ where $P_0$ is the saturated water vapor pressure, the data exhibits a very sharp adsorption/desorption of two water molecules per copper ion (referred to as "guest water" hereafter). In the following, the PCP fully occupied by guest water molecules is called **PCP-1**, while **PCP-2** refers to the one without guest water at low relative pressure. Heavy water adsorption/desorption isotherms were also measured (Fig. 1A), and we verified that the adsorption/desorption properties exhibit no isotope effects. We also performed thermogravimetric analyses (TGA) of **PCP-1** (Supplementary Information). The data revealed another water molecule (referred to as "coordinated water" hereafter) per copper ion that is strongly bound to the PCP framework (Fig. S2,3). As shown below, this corresponds to the water molecule coordinated to the copper ion.

**Characteristic of ices; ordered structure of guest water in PCP-1.** We performed single-crystal X-ray diffraction (SXRD) analysis of **PCP-1** to investigate whether guest water in **PCP-1** has an ordered structure as seen in ices or disordered structure as seen in liquids[12]. Fig. 1B shows the unit cell of **PCP-1**, which is a 2D sheet structure with a square lattice framework. The 2D sheets are stacked along the c-axis (Fig. 1C) with a separation distance of 6.9 Å, creating two kinds of quasi-1D channels (CH1 and CH2) with approximately 1 nm diameter. We observed all water molecules i.e., three per copper ion, in CH1, which means CH2 is empty (Fig. 1B). The fact that oxygen atoms of water molecules are observed by SXRD means that confined water molecules form an ordered structure along the channel (Fig. 1C). This is characteristic in ices[12], although measured at room temperature. Three types of water molecules (whose oxygen atoms are colored red, blue, and purple) exist in different environments. The red-colored oxygen atoms are in close proximity to the copper ions (green) and strongly bound by a coordination bond (coordinated water). In contrast, the other two oxygen atoms (blue and purple) are located near the center of the channel at different positions along the c-axis (sites I and II in Fig. 1C, respectively). The distances between blue (site I) and red oxygens, and purple (site II) and brown oxygens are 2.75(1) Å and 3.219(8) Å, respectively (Fig. 1B). This means that these are the guest water molecules weakly bound by H-bonding. The analyses are supported by powder crystal X-ray diffraction (PXRD) and simulation results (Fig. S4). Also, the distance between the four water molecules (2.87(1) Å) at sites I and II suggest the existence of H-bonding network among them (Fig. 1B) [16]. However, the guest water molecules at sites I and II are separated by a distance of 3.33(2) Å or 3.58(2) Å, which is comparable to the upper limit of the H-bond length[16]. This indicates considerably weak H-bond interactions between them (Fig. 1C) [16,17]. We also confirmed that there are no differences between the framework structures of **PCP-1** and **PCP-2** (Fig. S4). The averaged O-O distance among neighbored water molecules is calculated as (2.75+2.87+2.87+3.33+3.58)/5 = 3.08 Å. This corresponds to the density of 1.02 g/cm$^3$, and water density in **PCP-1** is comparable to liquid water: 0.997 g/cm$^3$ at 298 K, 0.1 MPa [17].

**Characteristic of liquids; significant fraction of broken H-bonds of confined water in PCP-1.** SXRD results show that the periodic array of the oxygen atoms of guest and coordinated water molecules are realized in CH1 even at room temperature. However, it is quite difficult to determine the positions of hydrogen atoms by SXRD measurements. To investigate the H-bond structure, we measured infrared (IR) absorption spectra of the OH stretching modes of confined water, which is sensitive to the H-bonds state [16,18,19]. The blue curve in Fig. 2A shows the IR absorption spectrum of **PCP-2** with only coordinated water. It shows two sharp absorption peaks (H1 and H2) owing to the OH stretching modes of coordinated water. This interpretation was confirmed by the isotopic effect observed for the two absorption peaks (see Supplementary Information). The IR spectrum of PCP-2 with coordinated D$_2$O (red curve in Fig. 2A) shows new absorption peaks (D1 and D2) at frequencies that are $2^{-1/2}$ times lower than those of H1 and H2, respectively [20]. As shown by the black curves in Fig. 2A, each peak is well-defined by the Lorentz function with parameters shown in Table 1, which means that the inhomogeneity of H-bonding structure of coordinated water is negligible. Due to the strong correlation between the resonance frequency of the OH stretching mode and H-bond length (O-O distance between water molecules), the peaks H1 (at 3450 cm$^{-1}$) and H2 (at 3655 cm$^{-1}$) can be assigned to H-bonded OH (HB-OH) with a H-bond length of ≈3 Å and non-H-bonded OH (Free-OH) [16], respectively. The number of H1 and H2 arms determined from the absorbance area normalized by oscillator strength are almost the same (see Supplementary Information), which means that every coordinated water molecule has both HB-OH and Free-OH arms. The crystallographic data (data file S1) shows that several oxygen atoms (colored brown in Fig. 2C) are located within ≈3 Å. However, only one atom satisfies the tetrahedral geometry formed by the sp$^3$ hybrid orbital of the oxygen atoms in coordinated water molecules. As shown in Fig. 2C, the HB-OH arm of coordinated water is H-bonded to an oxygen atom of an isophthalate in the adjacent layer, and the other OH arm is directed to the interior of the 1D nanochannel, which plays a role of guest water attractions. Furthermore, based on this knowledge and the SXRD result (Fig.1B), we can infer the H-bond network structure of the guest water as shown in Fig. 2D. Due to the negligible H-bonds between guest water in site I and II (Fig.1C), guest water molecules in site I have free or (considerably weak H-bonded) OH-arms.

The blue curve in Fig. 2B shows the IR absorption spectrum of **PCP-1** adsorbing guest water. A broad absorption band (H3) is observed in the frequency range of the OH stretching modes. H3 shows an approximately $2^{-1/2}$ times isotopic shift by replacing all the guest and coordinated H$_2$O with D$_2$O (red curve in Fig. 2B), indicating that H3 is due to the adsorbed H$_2$O. The broad absorption band H3/D3 is decomposed into three Gaussian functions and one Lorentz function (H4/D4) as shown in Fig. 2B. These are assigned to OH stretching modes with strong (3202 cm$^{-1}$), intermediate (3390 cm$^{-1}$), weak (3539 cm$^{-1}$), and no H-bond (H4, 3630 cm$^{-1}$) [19]. The surprising point is the existence of weak H-bonded and free OH, which is a characteristic of liquids not seen in ices[19,21]. Water in **PCP-1** has both characteristics of liquids (broken H-bonds) and ices (ordered structures). From the absorbance area of H4 peak normalized by oscillator strength, we can estimate the number of free-OH in **PCP-1**. Table 1 shows that the absorbance area of H4 peak is almost the same as that of H2 (Free-OH of coordinated water). Because of the small difference in peak position between H2 and H4, their oscillator strengths are comparable to each other[16]. Therefore, the number of free-OH in **PCP-1** is almost the same as that of coordinated water (four per unit cell). This is consistent with the inferred H-bonds structure in Fig.2D which shows the four free (or considerably weak H-bonded) OH-arms of guest water in site I per unit cell. The mean number of H-bonds per hydrogen atom in guest and coordinated water molecules is estimated as (24-4)/24=0.83, which lies between those of liquids (0.55[8]) and ices (≈ 1[12]). Interestingly, this indicates that despite the periodical array of oxygens that is characteristic in ices, confined water does not satisfy Pauling's ice rule which states that each water molecule offers/accepts two hydrogen atoms to/from two neighboring molecules and has four H-bonds [22].

We also investigated a dynamic property of confined water molecules by time-lapse IR spectroscopy. We prepared **PCP-1** adsorbing the mixture of H$_2$O and D$_2$O. We exposed the sample to D$_2$O gas vapor and measured the spectral changes in H3, which provides information about the diffusion dynamics of guest water molecules. The result is shown in Fig. 3A. D3 gradually increased in height with time (*t*), whereas H3 gradually decreased. These spectral changes are attributed to the exchange of guest H$_2$O/HDO with D$_2$O molecules outside the nanochannels. The H3 area in Fig.3A normalized by H3 area in Fig.2B provides the number of hydrogen atoms per unit cell, $n_{H\text{-}atom}(t)$, which is shown in Fig. 3B. It decreases exponentially (green curve), as expected from the 1D diffusion model [23],

$$n_{H\text{-}atom}(t) \propto \exp(-t/\tau) \qquad (1)$$

The time constant $\tau = 3.8 \times 10^3$ s determined from the fit allows us to deduce the diffusivity ($D$) of water in the nanochannel using the following relation [23],

$$\tau = l^2/(3D) + l/\alpha \qquad (2)$$

Here, $l$ and $\alpha$ are the lengths of the 1D nanochannel and surface permeability, respectively. The root-mean-square of the channel length $<l^2>^{1/2}$ is determined as 0.8 μm by using Scanning Electron Microscope (SEM) (Fig. S5). In general, the surface permeability influences the molecular exchanges with gas molecules in the case of small powder particles. It hinders the molecules from entering and leaving the pores. The surface permeability of **PCP-1** is unknown, but at least a positive number. Therefore, this allows us to determine the lower limit of the proton diffusivity as $0.6 \times 10^{-16}$ m$^2$/s, which is higher than that of ice VII ($2.6 \times 10^{-19}$ m$^2$/s at 300 K, 10 GPa [24]). Since the microscopic hopping process is induced by molecular rotations, this result shows that the structural rearrangements of confined water molecules occur faster in **PCP-1** than in solid ices at the same temperature.

**Discussion**

The experimental results clearly show the duality of the confined water in the 1 nm nanochannel of **PCP-1**; an ordered structure (characteristics in ices), a significant fraction of broken H-bonds $N_{broken\text{-}H}$ (17%) (characteristics in liquids), and large diffusivity. It is commonly believed that a solid-liquid critical point does not exist in the bulk phase diagram. This is because of the famous symmetry argument that an isotropic liquid cannot continuously transform into a solid with a discrete symmetry [14,15]. Therefore, in this sense, one can state that the confined water in the nanospace of **PCP-1** is exotic with no corresponding states in the bulk phase diagram.

We found that the structural and dynamical properties of water in **PCP-1** are quite similar to those of ***solid-liquid supercritical water*** in CNTs[12]. The molecular simulation study clearly showed that water in CNTs with 1 nm diameter has a solid-liquid critical point in the high pressure and ambient temperature region of the phase diagram. $N_{broken\text{-}H}$ changes discontinuously (from 0% to ~50%) during the solid-liquid phase transition that occurs below the critical point. Above the critical point, however, the molecular simulation showed that $N_{broken\text{-}H}$ changes continuously in between the limits of ices and liquid phases. Therefore, the intermediate value of $N_{broken\text{-}H}$ (17%) observed in **PCP-1** is characteristics of the solid-liquid supercritical water predicted in CNTs. Furthermore, the ordered square water (Fig. 1B) is the same structure as predicted in the supercritical region[12], which also supports the similarity. However, it should be noted that it remains unclear whether water in **PCP-1** is ***solid-liquid supercritical water***. Further experiments such as investigation of thermodynamic responses must be performed to clarify this point.

Finally, we comment on the origin of the exotic state. It is trivial that the ordered structure of oxygens in CH1 arises from the periodic array of the attractive potentials in **PCP-1**. Also, the existence of broken H-bonds (Fig.2D) probably comes from the structure of the PCP framework that aligns guest water molecules in site I and II with separation distances comparable to the upper limit of the H-bond length (Fig. 1C). This implies that attractive potentials of inner walls are strong enough to promote square water structure but not enough to freeze them on each site, resulting in both characteristics of ices and liquids.

In summary, we observed an exotic water in the 1 nm-size hydrophilic nanospaces of **PCP-1**. We found that the structural properties of water in **PCP-1** are quite similar to those of *solid-liquid supercritical water* in CNTs. The designed hydrophilic nanospaces of **PCP** will open up new experimental systems at ambient pressure to clarify the existence of the predicted solid-liquid critical points, which is one of the unresolved issues in physics. Also, nanoconfined water is related to the number of biological processes, such as protein folding, enzymatic reaction, and biological surface hydrations. Our results will contribute to a greater understanding of an important role of nanoconfined water in biological activities as well as to control chemical reactions using exotic water in PCPs.

## Methods

**$H_2O$ and $D_2O$ sorption isotherm for PCP-2 at 298 K.** The $D_2O$ and $H_2O$ sorption isotherms were obtained using a BELSORP-aqua3 instrument. All of the adsorbents were degassed via freeze-pump-thaw cycles using liquid $N_2$. In prior to sorption measurement, the powder crystals of **PCP-1** were placed in sample cells, which were evacuated at room temperature for more than 5 h and then purged with dry He to afford **PCP-2**. The sorption isotherm measurements were conducted, without exposing **PCP-2** to air, using an equilibrium time of 600 seconds for taking each equilibrium point. The sorption measurements were conducted up to relative pressures ($P/P_0$) of ca. 0.95 (Fig. 1A). The desorption isotherm of **PCP-1** for water vapor at 298 K is shown in Fig. 1A, in which the number of $H_2O$ molecule per Cu is calculated based on the amount of water adsorbed in **PCP-1**.

**Samples for IR measurements.** The sample was a pellet (~ 100 μm thickness) made by pressing microcrystalline powder sample of **PCP-1** (the typical size is 0.67μm). The pellet was placed in a chamber, wherein the humidity was regulated to control the adsorbed water amount in **PCP-1**. Chamber volume is $\pi \times 1.28$ cm$\times 1.28$ cm$\times 0.7$ cm $= 3.6$ cm$^3$. The low humidity below 4 % is achieved by flowing dry air into the chamber. The high humidity (~30 %) is obtained by the mixture of dry air and saturated steam. The humidity in the chamber is measured by means of a digital humidity meter (Model CTH-1100, Custom, Tokyo). The IR spectra were performed at ambient temperature by using Hyperion Fourier Transform Infrared (FTIR) microscope connected to the Bruker VERTEX 80v spectrometer at a spectral resolution of 4 cm$^{-1}$. The steady states of coordinated and guest water were measured by 64 accumulation scans. The used $D_2O$ gas molecules were evaporated from 99% concentration of $D_2O$ (CAS. No. 7789-20-0, Cambridge Isotope Laboratories Inc., product code DLM-4-99-1000).

**Isotopic substitutions.** Isotopic substitution of guest and coordinated $H_2O$ with $D_2O$ was achieved just by exposing **PCP-1** to $D_2O$ vapour at ~30 % RH. The concentration of deuterium atoms of guest and coordinated water is controlled by exposing **PCP-1** to the proper mixture of the deuterium oxide and distilled water (CAS. No. 7732-18-5, Wako Pure Chemical Industries Ltd., Osaka, product code 049-16787).

**Acknowledgments:**
S.K., H.S., N.H., and T.A. are thankful to the Japan Society for the Promotion of Science (JSPS) for KAKENHI Grant-in-Aid for Specially Promoted Research (25000007). S.K., H.S., and N.H. acknowledge the financial support of Scientific Research (S) (18H05262). N.H. acknowledges JSPS for a KAKENHI Grant-in-Aid for Young Scientists (B) (16K17959) and Scientific Research (B) (18H02072). K.T. is thankful to the Scientific Research (S) (17H06124).

**Author contributions:** T.I., T.A., S.K., and K.T. conceived the experiments. T.I. built the set-up, performed the experiments and analyzed the experimental data with contributions from T.A. and K.T. K.O., N.H., and H.S. synthesized and characterized the PCP sample. T.I., T.A., and K.T. developed the analytical model. All the authors contributed to discussing the results and writing the paper.

**Additional Information:**

**Accession codes:** All data necessary to support the conclusions are available in the manuscript or supplementary materials. The x-ray crystallographic coordinate for a structure reported in this paper have been deposited at the Cambridge Crystallographic Data Centre under deposition number CCDC 1893959.

**Competing interests:** Authors declare no competing interests.

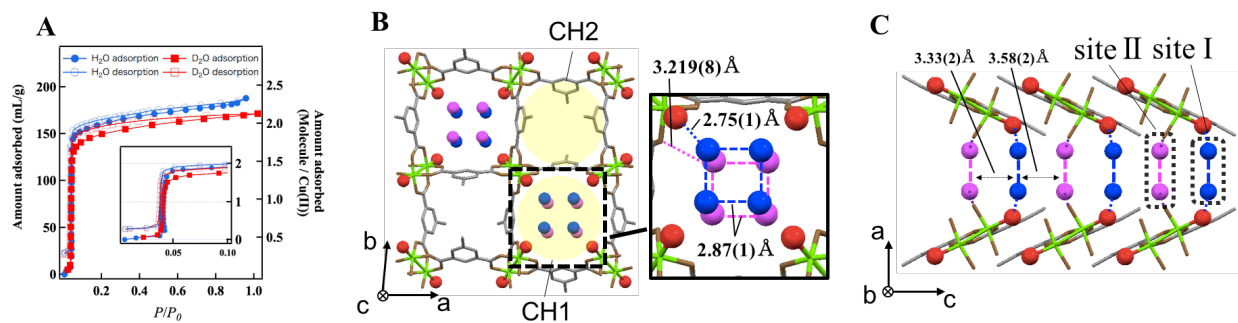

**Fig. 1. Water adsorbing PCP.** (A) Adsorption/desorption isotherm of water adsorbing PCP at 298 K. (B) The unit cell of the crystal structure of **PCP-1**. The bondings are shown, but atoms are colored as follows: Cu, green; C, gray; O of isophtalates, brown; O of coordinated water, red; O of guest water, blue and purple. Hydrogen atoms are omitted. The magnified view shows the confined water structure with distances between oxygen atoms. See also Fig. S1. (C) Side view of the channel CH1 of **PCP-1**.

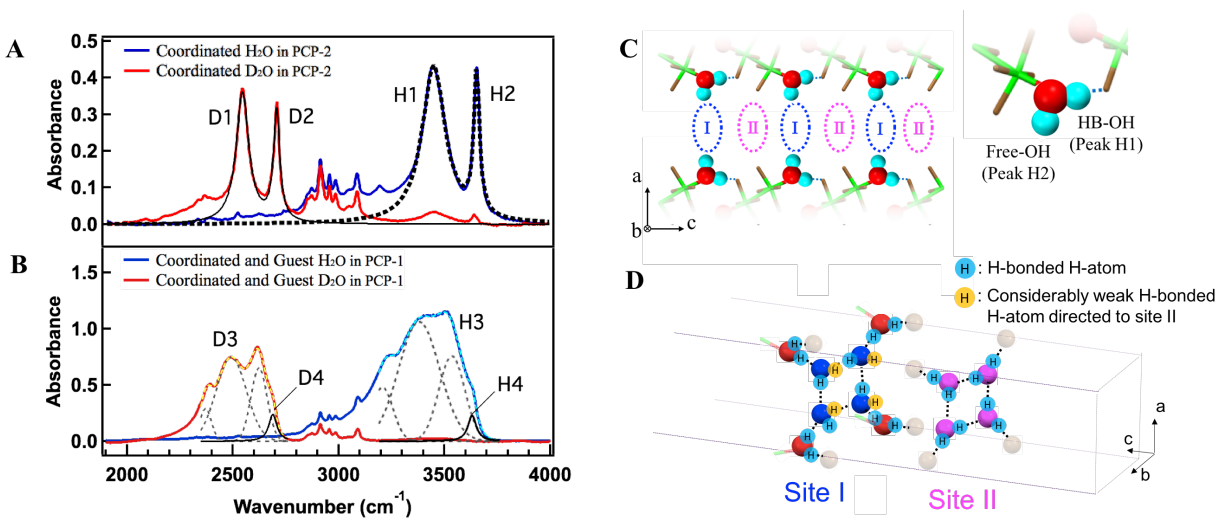

**Fig. 2. Hydrogen bond structure of coordinated and guest water.** (A) The IR absorption spectra of **PCP-2** adsorbing only coordinated $H_2O$ (blue line) and $D_2O$ (red line) at 295 K below 4 % RH. The small peaks (2800 ~ 3100 cm$^{-1}$) correspond to the CH stretching modes of the framework. The black broken and solid lines are the result of Lorentz curve fittings of absorption peaks due to coordinated $H_2O$ and $D_2O$, respectively. (B) The IR absorption spectra of **PCP-1** adsorbing $H_2O$ (blue line) and $D_2O$ (red line) at 30 % RH. Each IR band is well decomposed into three Gaussian functions (black dashed) and one Lorentz function (solid line). Yellow and cyan dashed lines are the sum of four deconvoluted functions of the red and blue line, respectively. (C) The schematic diagram of H-bond structure of coordinated water in **PCP-2**. Hydrogen atoms are colored by cyan. (D) The schematic diagram of H-bond structure of coordinated and guest water in **PCP-1** inferred from SXRD and IR experiment results.

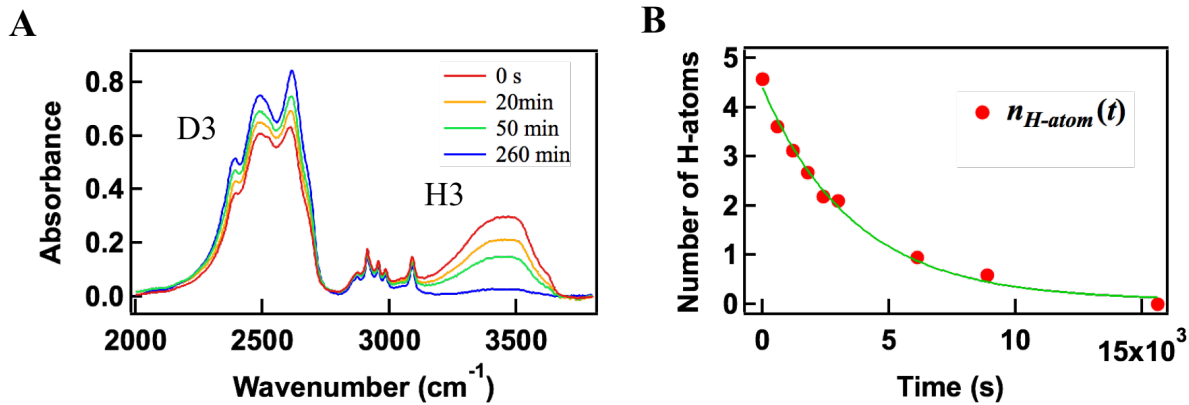

**Fig. 3. Dynamic properties of confined water.** (A) The series of IR absorption spectra after exposing to $D_2O$ vapor. The small peaks (2800 ~ 3100 cm$^{-1}$) corresponding to the CH stretching modes of the framework show no spectral changes. (B) The number of hydrogen atoms per unit cell ($n_{H\text{-}atom}(t)$). The green solid line shows a curve fitting result with an exponential function.

**Table 1**. **Results of peak positions, FWHMs, and ratio of absorbance area for Lorentz fitting of coordinated $H_2O$ and $D_2O$ in Fig. 2A and H4&D4 in Fig. 2B.**

|    | Peak position (cm$^{-1}$) | FWHM (cm$^{-1}$) | Ratio of absorbance area |
|----|---------------------------|------------------|--------------------------|
| H1 | 3450                      | 140              | 4.8                      |
| H2 | 3655                      | 33.5             | 1                        |
| H4 | 3630                      | 50.9             | 0.91                     |
| D1 | 2544                      | 73.5             | 2.1                      |
| D2 | 2706                      | 36.8             | 0.76                     |
| D4 | 2689                      | 50.8             | 0.85                     |